\documentclass[conference]{IEEEtran}
\IEEEoverridecommandlockouts
\usepackage{bbm}
\usepackage{url}
\usepackage{stfloats}
\usepackage{amsfonts}
\usepackage[dvips]{graphicx}
\usepackage{times}
\usepackage{cite}
\usepackage{amsmath}
\usepackage{array}
\usepackage{amssymb}

\usepackage{stfloats}
\usepackage{graphicx}
\usepackage{footnote}
\usepackage{booktabs}
\usepackage{array}
\usepackage{algorithm}
\usepackage{subeqnarray}
\usepackage{cases}
\usepackage{threeparttable}
\usepackage{color}
\usepackage{textcomp}

\usepackage{epstopdf}
\usepackage{algpseudocode}
\usepackage{bm}
\usepackage{multirow}
\usepackage[labelformat=simple]{subcaption}
\usepackage{adjustbox}
\newtheorem{proposition}{Proposition}

\usepackage{cite}
\usepackage{graphicx}

\def\BibTeX{{\rm B\kern-.05em{\sc i\kern-.025em b}\kern-.08em
    T\kern-.1667em\lower.7ex\hbox{E}\kern-.125emX}}

\begin{document}

%
% The "title" command has an optional parameter, allowing the author to define a "short title" to be used in page headers.
\title{Average-Case Analysis of Greedy Matching for D2D Resource Sharing}

\author{\IEEEauthorblockN{Shuqin Gao$^\ast$, Costas Courcoubetis$^\dagger$, and Lingjie Duan$^\ast$}
\IEEEauthorblockA{$^\ast$Engineering Systems and Design Pillar, Singapore University of Technology and Design\\
$^\dagger$School of Data Science, The Chinese University of Hong Kong, Shenzhen}
}

\maketitle


%
% By default, the full list of authors will be used in the page headers. Often, this list is too long, and will overlap
% other information printed in the page headers. This command allows the author to define a more concise list
% of authors' names for this purpose.

%
% The abstract is a short summary of the work to be presented in the article.
\begin{abstract}
Given the proximity of many wireless users and their diversity in consuming local resources (e.g., data-plans, computation and even energy resources), device-to-device (D2D) resource sharing is a promising approach towards realizing a sharing economy. In the resulting networked economy, $n$ users segment themselves into sellers and buyers that need to be efficiently matched locally. This paper adopts an easy-to-implement greedy matching algorithm with distributed fashion and only sub-linear $O(\log n)$ parallel complexity, which offers a great advantage compared to the optimal but computational-expensive centralized matching. But is it efficient compared to the optimal matching? Extensive simulations indicate that in a large number of practical cases the average loss is no more than $10\%$, a far better result than the $50\%$ loss bound in the worst case. However, there is no rigorous average-case analysis in the literature to back up such encouraging findings, which is a fundamental step towards supporting the practical use of greedy matching in D2D sharing. This paper is the first to present the rigorous average analysis of certain representative classes of graphs with random parameters, by proposing a new asymptotic methodology. For typical 2D grids with random matching weights we rigorously prove that our greedy algorithm performs better than $84.9\%$ of the optimal, while for typical Erd{\H{o}}s-R{\'e}nyi random graphs we prove a lower bound of $79\%$ when the graph is neither dense nor sparse. Finally, we use realistic data to show that our random graph models approximate well D2D sharing networks encountered in practice.

\end{abstract}


%
% Keywords. The author(s) should pick words that accurately describe the work being
% presented. Separate the keywords with commas.

%
% A "teaser" image appears between the author and affiliation information and the body 
% of the document, and typically spans the page. 

%
% This command processes the author and affiliation and title information and builds
% the first part of the formatted document.
\maketitle
\pagestyle{empty}

%%%%%%%%%%%%%%%%%%%%%%%%%%%%%%%%%%%%%%%%%%%%%%%%%%%%%%%%%%%%%%%%%%%%%%%%%%%%%%%%
\section{Introduction}

Thanks to advances in wireless and smartphone technologies, mobile users in proximity can use local wireless links (e.g., short-range communications) to share local resources (e.g., data-plans, computation and energy resources). For instance, subscribed users who have leftover data plans can set up personal/portable hotspots and share data connections to travelers with high roaming fees or those who face data deficit \cite{xuehe,data}. Besides sharing data-plans, mobile terminals with extra local memories can exchange the requested files with each other in the vicinity \cite{caching}. Furthermore, a user in need of bandwidth for video streaming can seek some neighboring users' assistance to download and forward video segments via local links \cite{microcast}. Given the large diversity for each user in the levels of her individual resource utilization, device-to-device (D2D) resource sharing is envisioned as a promising approach to pool resources and increase social welfare.

Some recent studies have been conducted for modelling and guiding D2D resource sharing in wireless networks (e.g., \cite{xuehe,data,microcast,D2Dcomputing,caching,energy}). As a node in the established D2D network graph, each mobile user can be a resource buyer or seller, depending on whether her local resource is sufficient or not. As in \cite{D2Dcomputing} and \cite{caching}, according to their locations, each user can only connect to a subset of users in the neighborhood through wireless connections, and the available wireless links are modelled as edges in the network graph. Sharing between any two connected users brings in certain benefit to the pair, which is modeled as a non-negative weight to the corresponding edge.

All these works (e.g., \cite{xuehe,data,microcast,D2Dcomputing,caching,energy}) optimize resource allocation by matching buyers and sellers in a centralized manner that requires global information and strict coordination. Hence the developed approaches cannot scale well in a scenario involving a large number of users, due to a large communication and computation overhead caused by the centralized nature of the proposed solutions. Carrying this argument further, the existing optimal weighted matching algorithms from the literature cannot be effectively used in the case of large user-defined networks due to their centralized nature and super-linear time complexity \cite{overview}. This motivates the need for developing distributed algorithms that exploit parallelism, have low computation complexity and good average performance for practical parameter distributions.

In the broader literature of distributed algorithm design for matching many buyers and sellers in a large graph,
a greedy matching algorithm of linear complexity is proposed in \cite{paper7,paper8} without requiring a central controller to gather all information. It simply selects each time the edges with local maximum weights and yields an approximation ratio of $1/2$ as compared to the optimum. A parallel algorithm is further proposed in \cite{logtime} to reduce complexity at the cost of obtaining a smaller approximation ratio than $1/2$. It should be noted that in the analysis of these algorithms, complexity and approximation ratio are always worst-case measures, but the worst-case approximation ratio rarely happens in most network cases in practice. Overall, there is a lack of average-case analysis of the greedy matching in the literature to compare its average performance with the optimal matching.

Since worst-case bounds no longer work for average-case analysis, we develop totally new techniques to analyze average performance. These techniques become more accurate when taking into account the structure of the network graph, and provide a very positive assessment of the greedy matching's average performance that is far from the worst case. Since the greedy matching can be naturally implemented in parallel by each node in the network, we also prove that with high probability (w.h.p.), the algorithm has sub-linear parallel complexity $O(\log n)$ in the number of users $n$. Our main contributions and key novelty are summarized as follows. 

\begin{itemize}
    \item \emph{New average-case analysis of greedy matching in 2D grid networks with random weights:} To provide rigorous results we first study the average performance of the greedy matching in the representative case of 2D grid network with random weights. In this case, dynamic programming cannot be directly applied since matching users does not divide the grid into sub-grids. We introduce a new asymptotic analysis method to prove that our greedy matching's average performance ratio as compared to the average of the optimal matching is at least $84.9\%$. If the greedy algorithm is allowed to run in parallel (as expected in practice), we prove that it has only sub-linear complexity $O(\log n)$ w.h.p. in 2D grids. Thus, our algorithm provides a great implementation advantage compared to the optimal but computational-expensive centralized matching.

    \item \emph{New average-case analysis of greedy matching in random networks:} We develop a new theoretic technique to analyze large Erd{\H{o}}s-R{\'e}nyi random graphs $G(n, p)$, where each of $n$ users connects to any other user with probability $p$. 
    For a dense random graph with constant $p$, we prove that the greedy matching achieves an average performance ratio that tends to $100\%$ as $n$ increases. 
    The analysis of sparse graphs is more challenging, yet we equivalently reduce to the asymptotic analysis of random trees. By exploiting the recursive nature of trees, we obtain rigorous average performance ratio bounds and parallel complexity $O(\log n)$ (w.h.p.) when $p<1/n$. We also prove that the average performance ratio reaches its minimum (still above $79\%$) when the graph is neither dense nor sparse.

    \item \emph{Application to practical scenarios:} We conduct experiments using real data for mobile user locations to simulate realistic D2D networks with given constraints on the maximum allowed communication distance between devices. We show that our analytical $G(n,p)$ performance measure approximates well many practical cases of such D2D sharing networks. To decide the maximum D2D sharing range among users, we take into account the D2D communication failure due to path-loss and mutual interference among matched pairs. The optimal sharing range is achieved by tradeoff between transacting with more devices but with the higher risk that the chosen best neighbor might not be effectively usable due to a communication failure.

%    Furthermore, after distributively paralleling the greedy matching at all the users, we also prove that the greedy algorithm is almost sure to run in $O(\log n)$ complexity once $p$ is not large.  

\end{itemize}

The paper is organized as follows. In Section \ref{sec:preliminaries}, we present our network model and the greedy matching algorithm for solving the D2D resource sharing problem in any network graph. In Section \ref{sec:grid}, we analyze the average performance ratios of the algorithm in the 2D grids. Sections \ref{sec:random} and \ref{sec:practical} extend the average-case analysis to random graphs and practical scenarios. Section \ref{sec:conclusions} concludes the paper.

\section{System Model and Preliminaries}\label{sec:preliminaries}

\subsection{System Model for D2D Resource Sharing}

We first describe our D2D resource sharing model that involves a large number of potential users to share resources with each other via local wireless links (e.g., short-range communications). In this model sharing takes place in repeated time intervals (or `rounds'). In each interval, we first run an algorithm to determine the pairs of users that are willing to exchange resources during the given round, and then realize the actual sharing of the corresponding resources as determined by the algorithm. The set of participating users and the available D2D links may be different for different rounds.

In each round, we form the `network graph' $G=(U,E)$, where $U$ is the set of nodes corresponding to users that are participating in the given round, and $E$ is the set of D2D links between participating users that are feasible to establish with some minimum level of performance (e.g., signal strength, actual physical distance, etc., depending on the application). For each user $u_i\in U=\{u_1,u_2,\cdots,u_n\}$, the subset $A(u_i)\subseteq U$ denotes the set of her neighbors in $G$, i.e., there is a feasible D2D link $e_{ij}\in E$ between $u_i$ and any $u_j\in A(u_i)$. Note that different definitions of `feasibility' for D2D links will imply a different set of edges $E$ between the users in $U$. Also the set $U$ is changing over time since new users may join the sharing economy and existing users may drop out after satisfying their needs or moving out of range.

With each edge $e_{ij}\in E$ there is an associated weight $w_{ij}\geq 0$ that models the (net) sharing benefit between this pair of users, if these users eventually transact with each other (i.e., are `matched' in our terminology), converted in some monetary basis (say $\$$). Let $W=\{w_{ij}\}$ be the weight vector over all edges of $G$. Note that our model is very flexible and can fit to various applications of D2D resource sharing by allowing for different ways to define the values for $w_{ij}$. For example, in a secondary data-plan trading market \cite{xuehe,data}, user $u_i$ with data-plan surplus shares her personal hotspot connection with neighboring user $u_j$ with high roaming fee, and weight $w_{ij}$ models the difference between user $u_i$'s saved roaming fee and the sharing cost (e.g., energy consumption in battery) of user $u_i$. In another example of cooperative video streaming \cite{microcast}, user $u_i$ seeks user $u_j$'s assistance to download video segments for local sharing, and $w_{ij}$ becomes the difference between the QoE improvement of user $u_i$ and the download cost of user $u_j$.  

\begin{figure}[t]\centering
\vspace{0cm}
\includegraphics[width=6.5cm]{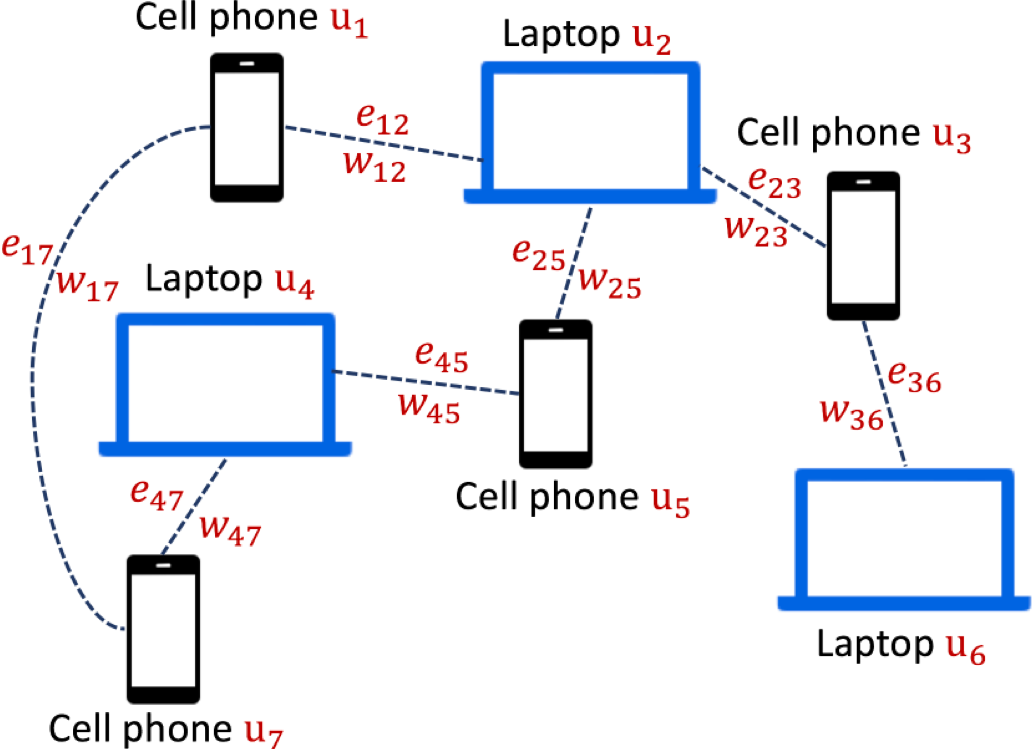}
\caption{An illustrative instance of the D2D resource sharing model with $n=7$ users is captured spatially.}
\label{fig:instance}
\vspace{-0.5cm}
\end{figure}

In any given round, our sharing model corresponds to an instance of a random weighted graph $(G=(U,E),W)$. A simple interpretation of the model is that a typical user, when participating, corresponds to a random node in $G$. In particular, we don't care for the actual identity of the participating users (after all, we care for the total value generated in the economy, summed over all participants). To simplify the model, we assume certain i.i.d. properties for the resulting stochastic process, i.e., in each round the set $U$ and the corresponding $E$, $W$ are i.i.d., with certain distributions. In particular, we assume that the weights $w_{ij}$ take values from a finite discrete set $V\hspace{-2pt}=\hspace{-2pt}\{v_1,v_2,\cdots,v_{K}\}$ according to the general probability distribution $Pr(w_{ij}\hspace{-2pt}=\hspace{-2pt}v_k)\hspace{-2pt}=p_k$ with $\sum_{k=1}^K p_k=1$.\footnote{We can similarly extend our average-case analysis to continuous weight distributions, though it is unlikely to have any practical interest.} Without loss of generality, we assume $0\hspace{-2pt}\leq \hspace{-2pt}v_1\hspace{-2pt}<\hspace{-2pt}v_2\hspace{-2pt}<\hspace{-2pt}\cdots\hspace{-2pt}<\hspace{-2pt}v_{K}$. A small-scale illustrative instance of the D2D resource sharing model is shown spatially on the ground in Fig.~\ref{fig:instance}, which can be abstracted to a weight graph $(G\hspace{-2pt}=\hspace{-2pt}(U\hspace{-2pt}=\hspace{-2pt}\{u_1,u_2,\cdots,u_7\},E\hspace{-2pt}=\hspace{-2pt}\{e_{12},e_{17},e_{23},e_{25},e_{36},e_{45},e_{47}\}),W\hspace{-2pt}=\hspace{-2pt}\{w_{12},w_{17},w_{23},w_{25},w_{36},w_{45}, w_{47}\})$.

In typical practices of D2D sharing (e.g., energy transfer), a user is only matched to a single neighbor (if any) to finally transact with. Keeping this simple but practical case of single matching per user\footnote{Allowing more concurrent matchings might not greatly improve performance, since
in practice most of the total benefit is usually obtained from one among the possible matchings (e.g., assume a Pareto distribution on the values of the possible matchings). Furthermore, obtaining the resources as a result of a single matching decreases the marginal benefit of the resources from more matchings.}, given a weighted graph $(G=(U,E),W)$, we would like to match the most valuable (with the highest value of $w_{ij}$) pairs of users to maximize the total sharing benefit (i.e., the `social welfare'). Assuming full and globally available information on $G$ and $W$, we formulate the social welfare maximization problem as a maximum weighted matching problem:
\begin{subequations}
\begin{align}
&\mathcal{P}_1:&\max \quad & \sum_{e_{ij}\in E} w_{ij}x_{ij}, \label{equ:objective}\\
&&\text {s.t.} \quad & \sum_{u_j\in A(u_i)} x_{ij}\leq 1, \quad \forall u_i\in U, \label{equ:matching1}\\
&&& x_{ij}\in \{0,1\},\quad\forall e_{ij}\in E, \label{equ:assignment}
\end{align}
\end{subequations}
where $x_{ij}$ is the binary optimization variable denoting whether edge $e_{ij}$ is included in the final matching ($x_{ij}=1$) or not ($x_{ij}=0$). Constraint \eqref{equ:matching1} tells that any user $u_i$ can only be matched to at most one user in her set of neighbors $A(u_i)$.

\subsection{Preliminaries of Greedy Algorithm}

According to \cite{overview}, to optimally solve the maximum weighted matching problem $\mathcal{P}_1$, one needs to centrally gather the weight and graph connectivity information beforehand. Further, searching for all possible matchings results in super-linear computation complexity, which is formidably high for a network with a large number $n$ of users. Alternatively, the greedy matching addresses these two issues by keeping information local and allowing the algorithm to be performed in a distributed fashion. Algorithm 1 outlines the key steps of the greedy matching algorithm (please see intuition in the text that follows).

\vspace{0.1cm}
\noindent\textbf{Algorithm 1:} Greedy matching algorithm for solving problem $\mathcal{P}_1$ for the graph ($G=(U,E), W$).

\textbf{Initialization:} $U'=U$; $A'(u_i) = A(u_i), \forall u_i\in U $; $x_{ij}$

$=0,\forall e_{ij}\in E$.

In each iteration, repeat the following two phases:

\textbf{Proposal phase:} 

For each unmatched user $u_i\in U'$:

\begin{itemize}
    \item User $u_i$ selects a user
    $u_{j^*}$ among her unmatched neighbors in $A'(u_i)$ with the maximum weight $w_{ij^*}$.
    \item User $u_i$ sends to $u_{j^*}$ a matching proposal.
\end{itemize}

\textbf{Matching phase:} 

For a user pair $(u_i,u_j)$ that both $u_i$ and $u_j$ receive proposals

from each other:

\begin{itemize}
    \item Match $u_i$ and $u_j$ by updating $x_{ij}=1$ and $U'=U'\setminus \{u_i,u_j\}$.
    \item Make $u_i$ and $u_j$ unavailable for matching with others, by updating $A'(u_{k})=A'(u_{k})\setminus \{u_i\}$ for any $u_{k}\in A'(u_i)$, and similarly for $u_j$.

\end{itemize}

Algorithm 1 is randomized\footnote{It randomizes in the selection of preferred neighbors in case there are multiple equally best choices in the proposal phase. 
A way to simplify this and make the algorithm deterministic is to assume that nodes are assigned unique IDs and that a node assigns priority in the case of ties to her neighbor with the highest ID value. This avoids loops and guarantees termination in $O(|E|)$ steps. In the rest of the paper we can assume this deterministic version for Algorithm 1.} and can be implemented \emph{distributedly}: at each time, each user uses local information to choose the unmatched neighbor with the highest weight as her potential matching partner; she will stop once this preference becomes reciprocal, or there are no available unmatched neighbors. This algorithm calculates a matching with total weight at least $1/2$ of the optimum (see \cite{paper8}). 
This worst-case approximation ratio of $1/2$ is achieved in a three-edge instance where the middle edge has slightly larger weight than its two adjacent edges, since the greedy matching chooses the middle edge while the optimal matching chooses the two side edges.

\subsection{Our Problem Statement for Average-Case Analysis}

Though the approximation ratio of Algorithm 1 is $1/2$ with half efficiency loss in the worst case, this ratio is achieved in the three-edge instance only when the middle edge has slightly larger weight than its two adjacent edges. In a large network instance, given the i.i.d. assumption regarding the choice of the weights,
it is improbable that the graph will consist of an infinite repetition of the above special weighted three-edge pattern which leads to the worst-case performance.
Hence, we expect the average performance ratio of the greedy matching to be much above $1/2$. We next provide the necessary definitions for our average performance analysis.

By taking expectation with respect to the weights in $W$ that are i.i.d. with a general discrete distribution $Pr(w_{ij}=v_k)=p_k,\forall k=1,\ldots,K$, we define the average performance ratio $PR(G)$ of Algorithm 1 for a given graph $G$ as follows:
\begin{align}
PR(G)=\frac{\mathbb{E}_{W}[\hat{f}(G,W)=\sum_{e_{ij}\in E} w_{ij}\hat{x}_{ij}]}{\mathbb{E}_{W}[f^\star(G,W)=\sum_{e_{ij}\in E} w_{ij} x^{\star}_{ij}]},\label{equ:average}
\end{align}
where $f^\star(G,W)$ and $\hat{f}(G,W)$ denote the total weights (i.e., social welfare) under the optimal matching and the greedy matching, respectively, $\{x^\star_{ij}\}, \{\hat{x}_{ij}\}$ being the corresponding matchings. Since over time the algorithm is repeated for new instances, the numerator and denominator correspond to the time-average of the social welfare obtained by running the greedy and the optimal algorithms, respectively. 

We next evaluate the performance ratio for several special forms of practical interest for $G$ that corroborate the excellent performance of the greedy matching, including 2D grids and the more general random graph $G(n,p)$ networks. In the case of random graphs, we must take expectation in \eqref{equ:average} over both $G$ and $W$. Besides, we will also prove the sub-linear computation complexity to run Algorithm 1 for these networks.

\section{Average-Case Analysis for D2D Sharing in 2D Grid Networks}\label{sec:grid}

In wireless networks, 2D grids are widely used to model social mobility of users (e.g., \cite{sensorgrid,gvgrid}). In this section, we analyze the average performance ratio and the parallel complexity of Algorithm 1 to validate its performance on planar user connectivity graphs. Note that the average-case analysis of 2D grids is an important benchmark for the more general random graphs analyzed in the following sections. 

\subsection{Average Performance Analysis of Optimal Matching}

It is impossible to obtain the exact value of the average total weight $\mathbb{E}_{W}[f^\star(G,W)]$ under the optimal matching due to the exponential number of the possible matchings. Instead, we propose a method to compute an upper bound for the denominator $\mathbb{E}_{W}[f^\star(G,W)]$ in \eqref{equ:average} using a methodology that holds for general graphs. This upper bound will be used to derive a lower bound for the average performance ratio $PR(G)$ in \eqref{equ:average} later.

In any graph $G=(U,E)$, each matched edge $e_{ij}\in E$ adds value $w_{ij}$ to the final matching. Equivalently, we can think of it as providing individual users $u_i$ and $u_j$ with equal benefit $w_{ij}/2$. For any user $u_i$, this individual benefit does not exceed half of the maximum weight of its neighboring edges. Using this idea and summing over all users, the total weight of the optimal matching is upper bounded by
\begin{align}
f^\star(G,W)\leq \frac{1}{2}\sum_{u_i\in U} \max_{u_j\in A(u_i)} w_{ij}. \label{equ:newoptimal}
\end{align}
By taking expectation over the weight distribution, we obtain the closed-form upper bound of the average total weight. 

\begin{proposition}\label{pro:generalupper}
For a general graph $G=(U,E)$ with the weight set $V\hspace{-2pt}=\hspace{-2pt}\{v_1,v_2,\hspace{-1pt}\cdots\hspace{-1pt},v_K\hspace{-1pt}\}$ and the weight distribution $P\hspace{-2pt}=\hspace{-2pt}\{p_1,p_2,\hspace{-1pt}\cdots\hspace{-1pt},p_K\hspace{-1pt}\}$, the average total weight of the optimal matching is upper bounded by
\begin{align}
\mathbb{E}_W\hspace{-1pt}[f^\star(\hspace{-1pt}G,\hspace{-2pt}W\hspace{-1pt})]\hspace{-2pt}\leq\hspace{-2pt} \frac{1}{2} \hspace{-2pt}\sum_{u_i\in U}\hspace{-2pt}\sum_{k=1}^K \hspace{-2pt}v_k((\sum_{i=1}^k p_i)^{|A(u_i)|}\hspace{-2pt}-\hspace{-2pt}(\sum_{i=1}^{k-1} p_i)^{|A(u_i)|}), \label{equ:generaloptimal}
\end{align}
where $|A(u_i)|$ is the cardinality of $A(u_i)$.
\end{proposition}

The proof is given in Appendix A.

\subsection{Average Performance Ratio of Algorithm 1}

\begin{figure}[t]\centering
\vspace{0cm}
\includegraphics[width=8cm]{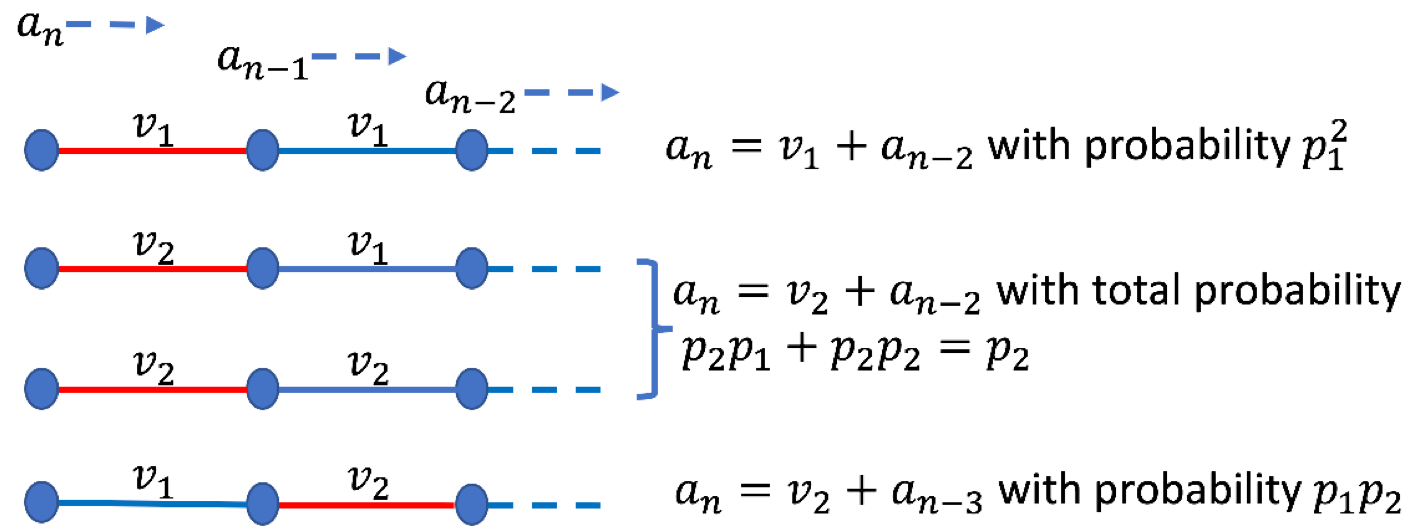}
\caption{Given the weight set size $K=2$ and $v_1<v_2$, an edge that certainly matches is always found in the first $2$ edges, as marked in red. There are totally $K^K=4$ weight combination cases of the first $2$ edges. In each case, after matching the red edge, the remaining graph is still linear but with smaller size $n-2$ or $n-3$.}
\label{fig:k2}
\vspace{-0.5cm}
\end{figure}

We next turn in the calculation of the average total weight of the greedy matching, i.e., the numerator $\mathbb{E}_{W}[\hat{f}(G,W)]$ in \eqref{equ:average}. In the case of 2D grid networks we cannot directly use dynamic programming since matching users does not divide the grid into sub-grids. Instead, we start by considering the simplest $1\times n$ grid where each user $u_i$ locally connects with two adjacent users $u_{i-1}$ and $u_{i+1}$ (except for starting and ending users $u_1$ and $u_n$). In such linear networks, for notational simplicity we use $e_i$ instead of $e_{i,i+1}$ to denote the connection between users $u_{i}$ and $u_{i+1}$, and similarly use weight $w_i$ instead of $w_{i, i+1}$. Without loss of generality, we assume that each user facing the same weights of the two adjacent edges assigns higher priority to match with the left-hand-side neighbor. This implies that Algorithm 1 becomes deterministic and returns a unique solution. 

We prove that given the weight set size $K$, an edge $e_i$ that has the local maximum weight (i.e., $w_i>w_{i-1}, w_i \geq w_{i+1}$ ) and will certainly match in Algorithm 1 can be found within the first $K$ edges of the linear network graph. Further, by considering all the $K^K$ weight combinations $\{w_1, \cdots, w_K\}$ of the first $K$ edges and the existence of edge $e_i$ that will certainly match, we derive the \emph{recursive formula} for $\{a_n\}$ where $a_n$ denotes the average total weight of the greedy matching in the linear graph with $n$ users. An illustrative example for $K=2$ is shown in Fig.~\ref{fig:k2} and the recursive formula averaged with respect to the probabilities is given by
\begin{align}
&a_n=p_1^2(v_1+a_{n-2})+p_2(v_2+a_{n-2})+p_1p_2(v_2+a_{n-3})\nonumber\\
&=p_1^2v_1+(p_2+p_1p_2)v_2+(p_2+p_1^2)a_{n-2}+p_1p_2a_{n-3}.\label{equ:greedyrecurrence}
\end{align}
Based on that, we derive $a_n=\frac{p_1^2v_1+(p_2+p_1p_2)v_2}{2p_2+2p_1^2+3p_1p_2}n+o(n)$ when $n$ is large by using asymptotic analysis. Similarly, for an arbitrary $K\geq 3$, we can also obtain the \emph{general formula} for the sequence $\{a_n\}$ as a function of $V=\{v_1,v_2,\cdots,v_K\}$ and $P=\{p_1,p_2,\cdots,p_K\}$.

\begin{figure}[t]\centering
\vspace{0cm}
\includegraphics[width=\hsize]{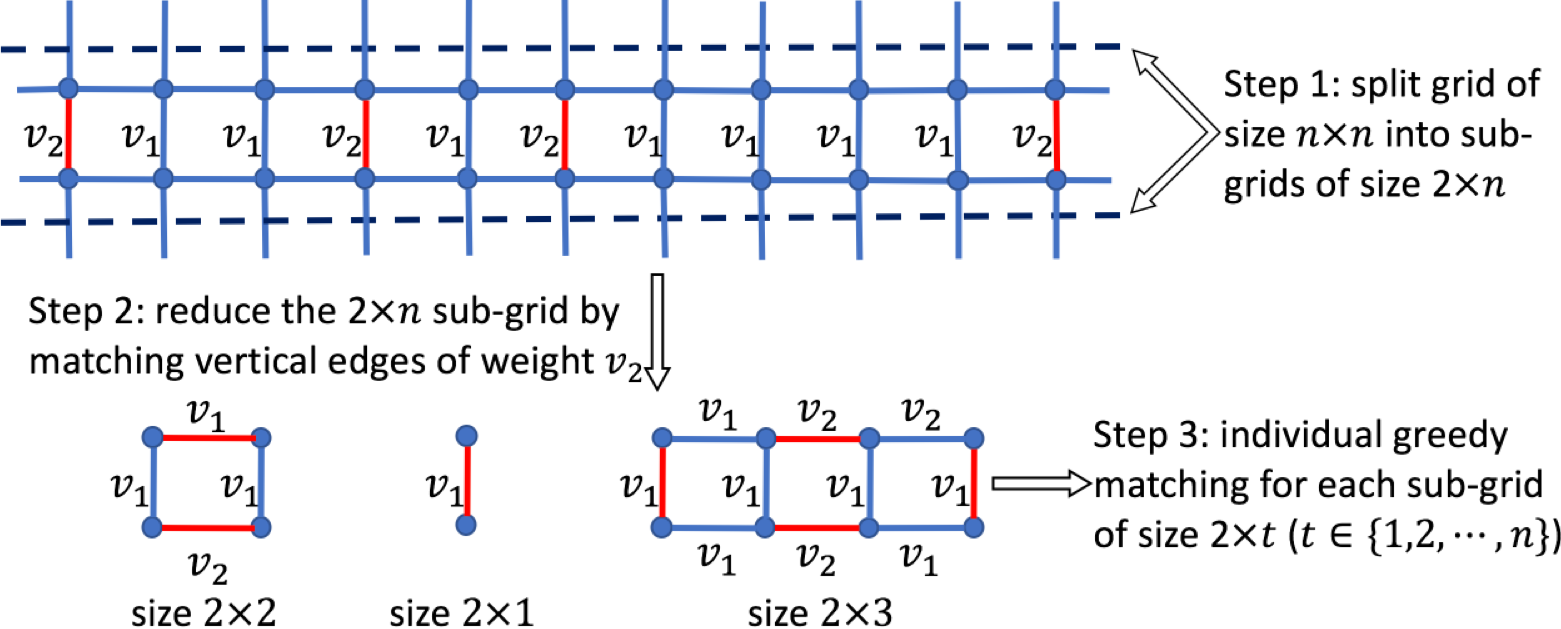}

\caption{Illustration of the graph decomposition process in three steps for analyzing greedy matching. Algorithm 1 adds the red-colored edges to the greedy matching in each step.}
\label{fig:grid}
\vspace{-0.5cm}
\end{figure}

Note that a simple extension of the previous result to $n\times n$ grid by dividing it into $n$ sub-grids of size $1\times n$ provides a bad lower bound because all the vertical edges become unavailable to match. Instead, we split the grid network into sub-grids in a way that keeps half of the vertical edges, and then estimate a lower bound of the average total weight. For illustration purpose, we treat the case of weight set size $K=2$, $v_1<v_2$. Our procedure involves the following three steps (as illustrated in Fig.~\ref{fig:grid}), and it is easy for the reader to generalize for $K>2$.

\vspace{0.1cm}

\noindent\emph{Step 1}: Split the $n\times n$ grid into $n/2$ sub-grids of size $2\times n$ by eliminating the corresponding edges.

\vspace{0.1cm}
\noindent\emph{Step 2}: For each $2\times n$ sub-grid from step 1, greedily add the vertical edges of (largest) weight value $v_2$ to the matching and remove the corresponding users. This results in further disconnecting the given sub-grid into smaller sub-grids of size $2\times t$, for appropriate values of $t$. Note that each such sub-grid only has vertical edges of (smaller) weight $v_1$.

\vspace{0.1cm}
\noindent\emph{Step 3}: For each sub-grid of size $2\times t$ from step 2, if $t=1$, simply add this single-edge to the matching. If $t>1$, greedily match the horizontal edges in the corresponding upper and lower linear networks first and remove the corresponding users; then match the remaining vertical edges of smaller weight $v_1$.

The above procedure allows us to find a lower bound on the matching obtained by Algorithm 1. This is obtained by i) eliminating from the set of edges that may be matched the vertical edges in step 1 (see Fig.~\ref{fig:grid}), ii) analyzing the greedy matching of the two horizontal linear networks in the resulting $2\times t$ sub-grids (using similar analysis as in \eqref{equ:greedyrecurrence}), and leaving out the possible matching of the rest vertical edges in step 3. By combining this lower bound for the performance of Algorithm 1 with the upper bound for the optimal matching in \eqref{equ:generaloptimal} we obtain the following bound for the average performance ratio.

\begin{proposition}\label{pro:gridbound}
In $n\times n$ grids with the weight set $V=\{v_1=1,v_2=2\}$ and uniform weight distribution $p_1=p_2=\frac{1}{2}$, the average performance ratio of Algorithm 1 satisfies $ PR(G)\geq 84.9\%$ when $n\rightarrow \infty$.
\end{proposition}

The proof is given in Appendix B. Observe that here we consider the case of weight set $V=\{1,2\}$ (`low' and `high') and uniform weight distribution $P=\{\frac{1}{2},\frac{1}{2}\}$ for illustration purpose. The analysis can be extended for any possible $V$ and $P$.

\subsection{Parallel Complexity Analysis of Algorithm 1}

In this subsection, we focus on analyzing the parallel running time of Algorithm 1, where a unit of time corresponds to one iteration of the steps of Algorithm 1. Similarly, we start with the simplest $1\times n$ grid. Let $H(u_i)$ denote the length of the longest chain (sequence of edges) that has non-decreasing weights and starts from $u_i$ towards the left or right side. Suppose that $w_{i-1}\leq w_{i-2}\leq\cdots\leq w_{i-H(u_i)+1}\leq w_{i-H(u_i)} >w_{i-H(u_i)-1}$ is such a longest chain. We claim that $u_i$ will terminate running Algorithm 1 (i.e., by being matched or knowing that there is no unmatched neighbors) within $H(u_i)/2$ time. This is easy to see since starting from time 0, the edge $e_{i-H(u_i)}$ will be included in the total matching in iteration 1, $e_{i-H(u_i)+2}$ in iteration 2, etc. Hence, in less than $H(u_i)/2$ steps, all neighbors of $u_i$ will have resolved their possible preferences towards users different than $u_i$, and subsequently $u_i$ will either be matched with one of her neighbors or be left with an empty unmatched neighbor set. As Algorithm 1 terminates when all users make their final decision, if the probability of any user in $G$ having a chain longer than $c\log n$ (i.e., $\max_{u_i\in U} H(u_i)> c\log n$) for some constant $c$ is very small, then the parallel execution of Algorithm 1 will terminate within $O(\log n)$ time with very high probability. 

Then, we extend our analysis to the $n\times n$ grid. Note that the number of possible chains that start from any given node $u_i$ and have non-decreasing weights is no longer two (toward left or right) as in $1\times n$ grid networks, but exponential in the size of the chain (since from each node there are $4-1=3$ `out' ways for the chain to continue), and such chains now form with non-negligible probability. This problem is not an issue for Algorithm 1 since every node will need to use priorities over ties among neighbors whose edge has the same weight. This significantly reduces the number of possible chains that are relevant to a user's decisions and we can prove the following proposition.
\begin{proposition}
In $n\times n$ grids, Algorithm 1 runs in $O(\log n)$ time w.h.p..
\end{proposition}

The proof is given in Appendix C. In conclusion, our distributed matching algorithm has low complexity and provides a great implementation advantage compared to the optimal but computational-expensive centralized matching.

\section{Average-Case Analysis for D2D Sharing in $G(n,p)$ Networks} \label{sec:random}

In practice, a mobile user may encounter a random number of neighbors. In this section, we extend our analysis to random networks $G(n, p)$, where $n$ users connect with each other with probability $p$ and hence each user has in the average (an order of magnitude) $d=np$ neighbors. Although the actual spatial distribution of users is not necessarily planar, such random graphs can still represent their connectivity on the ground and the analysis also holds.

We study the average performance ratio of Algorithm 1 in the cases of dense random graphs with a constant $p$ (i.e., dense since $d=np$ increases linearly in $n$) \cite{gnpbook2}, and sparse random graphs with a constant average neighbor number $d<1$ (i.e., $p<1/n$) \cite{gnpproperty}. Unlike the 2D grid networks, the structure of the random network $G(n, p)$ is no longer fixed due to the random connectivity. Though it is more technically difficult to analyze the average performance of Algorithm 1 for random graph structure, we are able to derive the ratio using statistical analysis in the two important cases below. For intermediate values of $d$ where our techniques cannot be applied, we have used exhaustive sets of simulations.

\subsection{Average-Case Analysis of Dense Random Graphs}

Given $p$ remains a constant, as $n$ increases, 
each user will have an increasing number of neighbors with the largest possible weight value $v_K$. Since such edges are preferred by greedy matching, as $n$ goes to infinity, the greedy matching will almost surely provide 
the highest possible total matching value of $nv_K/2$ ($n/2$ pairs of users with weight value $v_K$).

\begin{proposition}\label{pro:performancegnp}
For a random graph $G(n,p)$ with a constant $p$, the average performance ratio of Algorithm 1 satisfies $PR=100\%$ w.h.p..
\end{proposition}

The proof is given in Appendix D. In this result, in the definition of the average performance ration $PR$ we have taken expectation over both $G$ and $W$. Note that the computation complexity is not anymore $O(\log n)$ in this case due to the increasing graph density. An obvious bound is $O(|E|)$ proved in \cite{paper8}..

\subsection{Average-Case Analysis of Sparse Random Graphs}

In this subsection, we consider that the connection probability is $p=d/n$ and hence each user has a constant average number of neighbors $d(n-1)/n\to d$ as $n$ becomes large. We first prove low parallel complexity for Algorithm 1 as long as each user has a small enough number of neighbors to pair with
that depends on the distribution of the edge weights.

\begin{proposition}\label{pro:complexitygnp}
For $G(n,d/n)$ type of networks, Algorithm 1 runs in $O(\log n)$ time w.h.p. if $d<2/\max\{p_1,p_2,\cdots,p_K\}$.
\end{proposition} 

The proof is given in Appendix E. Note that this condition is always satisfied when $d<1$ as the weight probability $p_k\leq 1$ for any $k$. 

Next, we focus on studying the average performance ratio $PR$ for sparse random graphs $G(n,d/n)$. 
The average total weight of the optimal matching can be upper bounded by \eqref{equ:generaloptimal}, which works for any graph. Then, we only need to study the average total weight $\mathbb{E}_{G\sim G(n,d/n),W}[\hat{f}(G,W)]$ of the greedy matching. Note that when matching any graph $G$, we can equivalently view that the weight of any matched edge is equally split and allocated to its two end-nodes. Then we can rewrite the above expression as follows:
\begin{align}\label{equ:gnpsum}
\mathbb{E}_{G\hspace{-1pt}\sim G(n,\hspace{-1pt}d/n),W} [\hat{f}(G\hspace{-1pt},\hspace{-1pt}W)]\hspace{-1pt}=\hspace{-1pt}n\mathbb{E}_{G\hspace{-1pt}\sim G(n,d/n),\hspace{-1pt}W} [x_i(G\hspace{-1pt},\hspace{-1pt}W)],
\end{align}
where $x_i(G,W)$ is half of the weight of the matched edge corresponding to each user $u_i$ under the greedy matching.

We cannot use dynamic programming directly to compute the average weight $\mathbb{E}_{G\sim G(n,d/n),W} [x_i(G,W)]$ per user in \eqref{equ:gnpsum} since $G(n,d/n)$ may have loops and it cannot be divided into independent sub-graphs. 
Given that $n$ is large and assuming $d<1$, then graph $G(n,d/n)$, with very high probability, is composed by a large number of random trees without forming loops. In this case the matching weight $x_i(G,W)$ of user $u_i$ only depends on the connectivity with other users in the same tree. 
To analyze $x_i(G,W)$, we want to mathematically characterize such trees which turn out to be `small' because $d<1$. Note that, in $G(n,d/n)$, each user has $n-1$ independent potential neighbors, and its random neighbor number follows a binomial distribution $B((n-1),d/n)$ with mean $(n-1)d/n\rightarrow d$, as $n$ becomes large. This binomial distribution can be well approximated by the Poisson distribution $Poi(d)$ (with mean $d$). 
We define $T(d)$ as a random tree where each node in the tree gives birth to children randomly according to the Poisson distribution $Poi(d)$.
\begin{proposition}\label{pro:gnptree}
Given a sparse random network $G(n,d/n)$ with $d<1$ and sufficiently large $n$, 
the average matching weight of any node $u_i$ is well approximated by the average matching weight of the root node of a random tree $T(d)$, i.e.,
\begin{align}
\lim_{n\to \infty} \hspace{-3pt}\mathbb{E}_{G\hspace{-1pt}\sim G(n,\frac{d}{n}),W} [x_i(G\hspace{-1pt},\hspace{-2pt}W)]\hspace{-2pt}=\hspace{-2pt} \mathbb{E}_{T\hspace{-1pt}\sim T(d),W}[x_{root}(T\hspace{-1pt},\hspace{-2pt}W)].\label{equ:gnptree} 
\end{align}
\end{proposition}

The proof is given in Appendix F. We will show numerically later that the approximation in \eqref{equ:gnptree} yields trivial performance gap and remains accurate as long as $d\leq 10$. By substituting \eqref{equ:gnptree} into \eqref{equ:gnpsum}, we obtain approximately the average total weight $\mathbb{E}_{G\sim G(n,d/n),W}[\hat{f}(G,W)]$. Hence, it remains to derive the form of $\mathbb{E}_{T\sim T(d),W} [x_{root}(T,W)]$. Given the \emph{recursive nature} of trees, we are able to use \emph{dynamic programming} techniques.

The root node may receive multiple proposals from its children corresponding to different possible edge weights in the set $\{v_1,v_2,\cdots, v_K\}$, 
and will match to the one (of them) with the maximum weight. We define $y_k$, $k\in\{1,2,\cdots,K\}$, to denote the probability that the root node receive a proposal from a child who connects to it with an edge of weight $v_k$. Then, by considering all the possible weight combinations of the root's children, we can compute the probability to match a child with any given weight, using the proposal probabilities $y_k$. In a random tree $T(d)$, given the root node is matched with one of its children, the remaining graph can be divided into several sub-trees which are generated from the grand-child or child nodes of the root node. In any case, a sub-tree starting with any given node has the similar graph structure and statistical property as the original tree $T(d)$. 
Thus, we are able to analytically derive the recursive equations for finding the proposal probabilities $\{y_k\}$ for the root node. 

\begin{proposition}\label{pro:gnptreeyk}
In the random tree $T(d)$, for any $k\in \{1,2,\cdots,K\}$, the proposal probability $y_k$ from a child of edge weight $v_k$ to the root node is the unique solution to the following equation:
\begin{align}
y_k\hspace{-1pt}=\hspace{-1pt}e^{-\hspace{-1pt}(p_K\hspace{-1pt}+\hspace{-1pt}\sum_{j=k+1}^K y_jp_j) d}\sum_{i=0}^\infty \frac{(p_kd)^i(1\hspace{-1pt}-\hspace{-1pt}(1\hspace{-1pt}-\hspace{-1pt}y_k)^{i+1})}{(i+1)!y_k}.\label{equ:yk}
\end{align}
\end{proposition}

The proof is given in Appendix G. Though not in closed-form, we can easily solve \eqref{equ:yk} using bisection, and then compute the probability that the root node matches to a child with any given weight. Based on that we derive the average matching weight $\mathbb{E}_{T\sim T(d),W} [x_{root}(T,W)]$ of the root for \eqref{equ:gnptree} and thus $\mathbb{E}_{G\sim G(n,d/n),W}[\hat{f}(G,W)]$ in \eqref{equ:gnpsum}. Finally, by comparing with \eqref{equ:generaloptimal} under the optimal matching, we can obtain the average performance ratio of Algorithm 1.

\subsection{Numerical Results for Random Graphs}

\begin{figure}[t]\centering
\vspace{0cm}
\includegraphics[width=7cm]{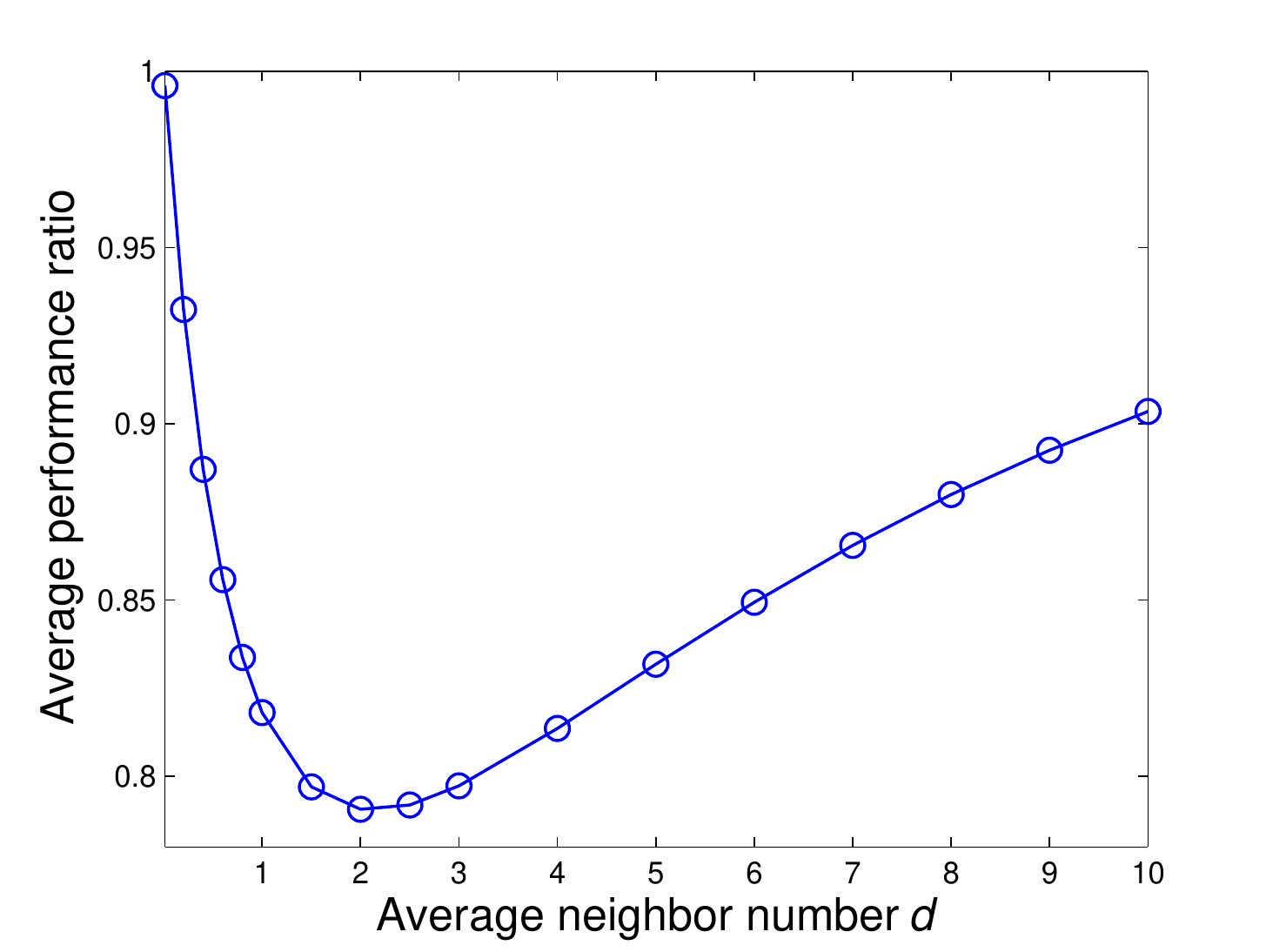}

\caption{The average performance ratio of Algorithm 1 in the large random graph $G(n,p=d/n)$ under different values of average neighbor number $d$.}
\label{fig:ratiognp}
\vspace{-0.5cm}
\end{figure}

Next, we conduct numerical analysis for sparse random graphs with $d<1$ and random graphs with finite $d\geq 1$. 
To do that by using analytic formulas, we need to approximate the random graph by random trees, and one may wonder if the approximation error is significant (when $d>1$). 
To answer this question, we consider large network size of $n = 10,000$, with edge weights uniformly chosen from the weight set $V=\{1,2\}$ (`low' and `high'). 
Our extensive numerical results show that the difference between the simulated average matching weight $\mathbb{E}_{G\sim G(n,\frac{d}{n}),W} [x_i(G,W)]$ and the analytically derived average matching weight $\mathbb{E}_{T\sim T(d),W} [x_{root}(T,W)]$ in the approximated tree $T(d)$ is always less than $0.05\%$ when $d<1$ and is still less than $1\%$ even for large $1\leq d\leq 10$. This is consistent with Proposition \ref{pro:gnptree}.

Fig.~\ref{fig:ratiognp} shows the average performance ratio of Algorithm 1, which is greater than $79\%$ for any $d$ value. 
It approaches $100\%$ as $d$ is small in the sparse random graph regime. Intuitively, when the average neighbor number $d$ is small and users are sparsely connected, both Algorithm 1 and the optimal algorithm try to match as many existing pairs as possible, resulting in trivial performance gap. 
When $d$ is large, each user has many neighbors and choosing the second or third best matching in the greedy matching is also close to the optimum. This is consistent with Proposition \ref{pro:performancegnp} for dense random graphs.

\section{Practical Application Aspects}\label{sec:practical}
In practice, the network graphs that one may obtain by restricting the D2D sharing range may have different distributions than the 2D grids and the $G(n,p)$ graphs used in our analysis. In addition to that, the actual performance of the algorithm might be degraded because of communication failures of nodes that are far or mutual interference among pairs. In this section, we provide an initial investigation of the above issues. We construct instances of a network graph based on realistic user location data and check how well our analytical $G(n,p=d/n)$ performance measure captures the actual performance of the greedy algorithm on the above graph instances by tuning $d$ to match the average number of neighbors in the realistic graph. Next, we analyze the impact of D2D communication failures on the optimum selection of D2D maximum sharing range.

\subsection{Approximating Realistic Networks}

\begin{figure}[t]\centering
\vspace{0cm}
\includegraphics[width=7cm]{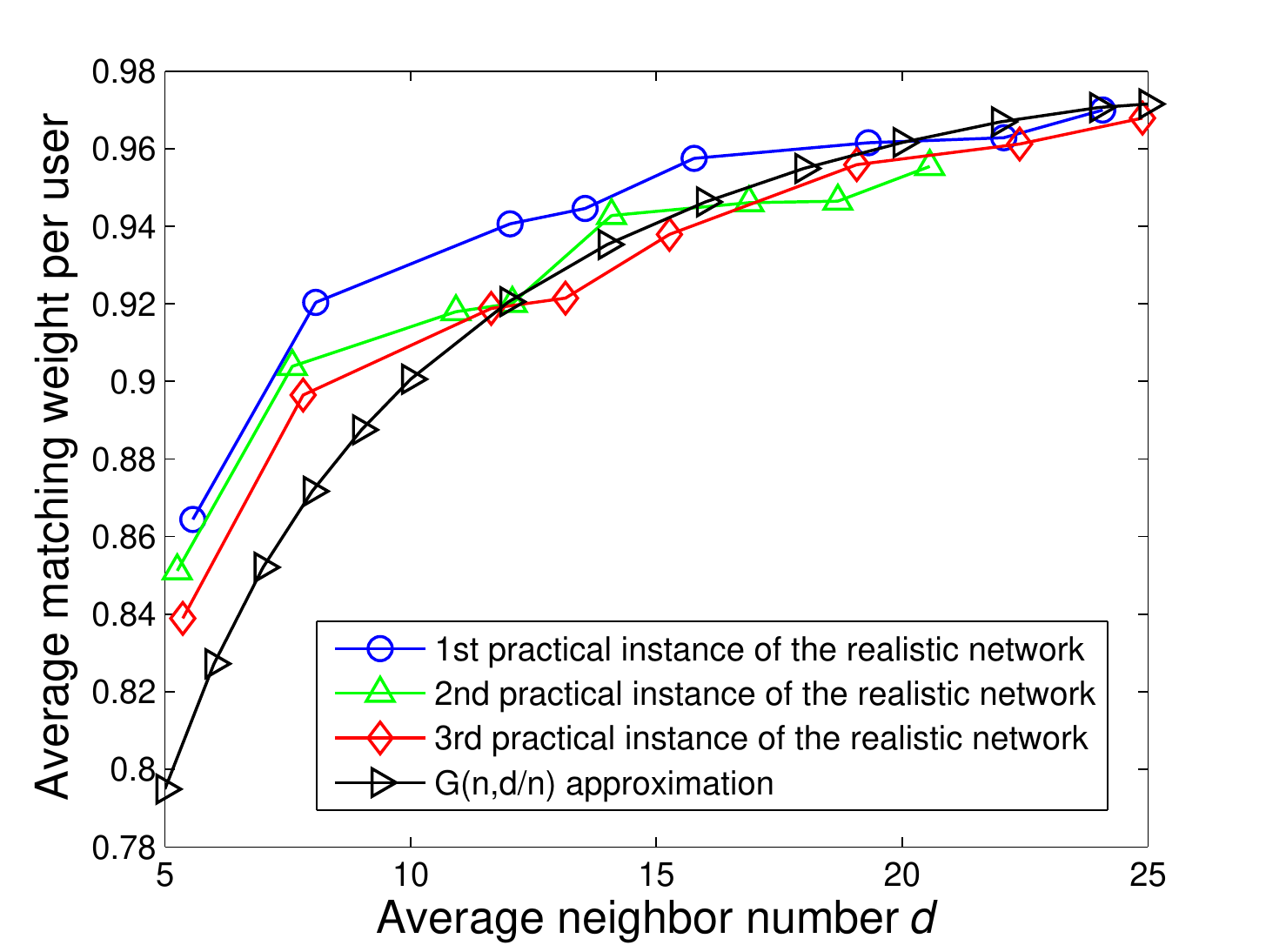}
\caption{The average matching weight per user of the greedy matching obtained by Algorithm 1 versus the average neighbor number $d$ in the three practical instances of the realistic network and the $G(n,p=d/n)$ network.}
\label{fig:realdataset2}
\vspace{-0.5cm}
\end{figure}

The $G(n,d/n)$ network studied in Section \ref{sec:random} assumes users connect with each other with the same probability $p=d/n$, and hence the average performance of Algorithm 1 in $G(n,d/n)$ is characterized by the average neighbor number $d$. However, in practice, the connectivity distribution of users can follow different laws due to the structure of the environment and the D2D communication limitations. To validate our analysis on real scenarios, we run our algorithm on graphs corresponding to realistic mobile user data and compare the numerical results with our analytically derived results for $G(n,d/n)$ using Propositions \ref{pro:gnptree} and \ref{pro:gnptreeyk}.

We use the real dataset in \cite{realdata} that records users' position information in a three-story university building. We choose three instances in the peak (in terms of density) time from the dataset and each instance contains hundreds of users. For these users, we assume that any two of them can share resources with each other when they are in the same story and the distance between them is less than the D2D sharing range $L$. By setting different values for $L$ the structure of the graph changes and the average number $d(L)$ of neighbors per node increases. In Fig.~\ref{fig:realdataset2}, we show the average matching weight (per user) of the greedy matching versus the average neighbor number $d$ (instead of versus $L$) for the three practical instances of the realistic user network and its $G(n,d/n)$ approximation. We observe that the average matching weight increases in $d$ since increasing $d$ (by increasing $L$) provides more sharing choices for each user and our performance measure obtained for $G(n,d/n)$ approximates well the performance for the realistic network, especially for large $d$.

\subsection{D2D Sharing Range under Communication Failures}

Our numerical results from the previous section suggest that, as expected, 
the average matching weight of the greedy matching increases with the D2D sharing range $L$. We wonder whether a large $L$ always benefits resource sharing. In fact, for two users who are connected and share resources via a D2D wireless link, a communication failure may occur due to the long-distance transmission or the mutual interference among different matched pairs. In our experiment, we consider that the resource sharing transmission between any two users fails with a probability that increases in the distance between them based on a practical path-loss model and the number of interfered pairs in proximity \cite{pathloss}. In our simulation experiment, we consider a large number $n=10,000$ of users uniformly distributed in a circular ground cell with radius of $R=1000$ meters and we can adjust the maximum sharing range $L$ inside which two users can apply D2D resource sharing.

\begin{figure}[t]\centering
\vspace{0cm}
\includegraphics[width=7cm]{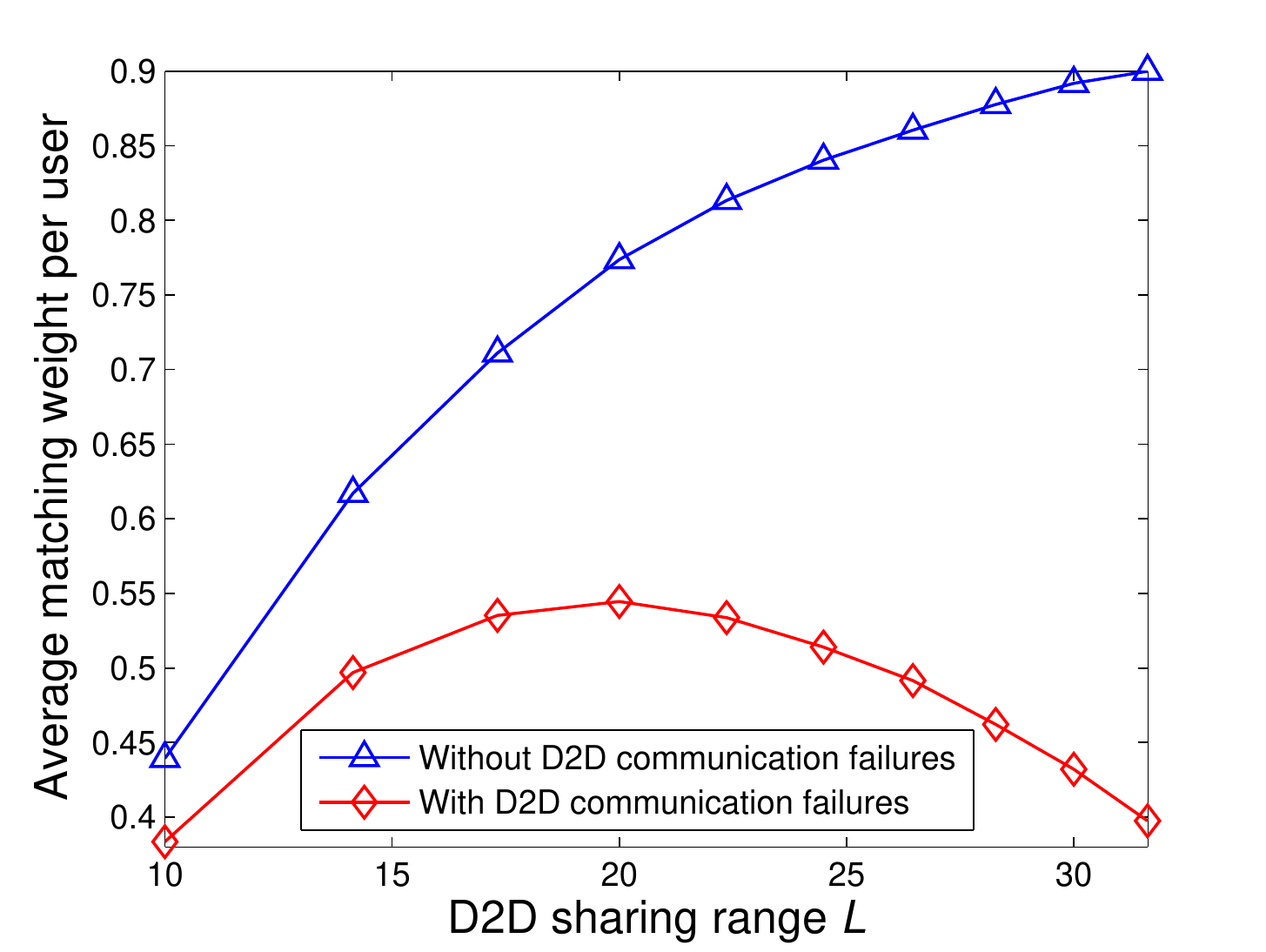}
\caption{The average matching weight per user of the greedy matching obtained by Algorithm 1 versus the maximum D2D sharing range $L$ with and without considering D2D communication failures. }
\label{fig:greedy_real}
\vspace{-0.5cm}
\end{figure}

In Fig.~\ref{fig:greedy_real}, we depict the average matching weight (per user) of the greedy matching versus the maximum D2D sharing range $L$ for two cases depending on whether or not we consider communication failures. Under communication failures, this weight first increases and then decreases. Intuitively, when the sharing range $L$ is small, each user has few potential users to share resources or interfere and failures occur rarely to be an issue. But when $L$ is large, since most of the neighbors are located remotely and the 
channels between matched pairs may cross each other, there is a high chance for the algorithm to choose such a remote neighbor to incur large path loss or interference. Then, if communication fails, this results in adding zero value to the total matching. This is in contrast to the case without failures, where the performance of the system is always increasing in $L$.

\section{Conclusions}\label{sec:conclusions}
In this paper, we adopt a greedy matching algorithm to maximize the total sharing benefit in large D2D resource sharing networks. 
This algorithm is fully distributed and has sub-linear complexity $O(\log n)$ in the number of users $n$. 
Though the approximation ratio of this algorithm is $1/2$ (a worst-case result), 
we conduct average-case analysis to rigorously prove that this algorithm provides a much better average performance ratio compared to the optimal in 2D grids and random networks $G(n,p)$ when these are sparse or very dense. We also use realistic data to show that our analytical $G(n,p)$ performance measure approximates well D2D networks encountered in practice. Finally, we consider the communication failures due to the practical implementation issues and study the best maximum D2D sharing range.
%In linear networks, to compute this ratio, we propose a graph separation method to obtain the upper bound of the optimal matching, and analytically derive the recurrence formula for the greedy matching. 
%We observe that the minimum value of the ratio is always achieved when all the possible weights are close to each other. 
%Further, we implement the similar average-case analysis of the algorithm in 2D grid models with the help of a new method that can compute upper bound of the optimal matching for any general graph. Finally, we extend our analysis to the more random network $G(n,p)$ and show that the greedy matching algorithm performs best when the network is extremely sparse or dense. 

\begin{appendices}
\section{Proof of Proposition 1}
For any user $u_i\in U$ with degree $d(u_i)$, the probability that the maximum weight of all the $d(u_i)$ neighboring edges is equal to or less than $v_k$ is given by $(\sum_{i=1}^k p_k)^{d(u_i)}$ where $\sum_{i=1}^k p_k$ is the probability that one edge has weight equal to or less than $v_k$. Thus, the average maximum weight of all $d(u_i)$ edges incident to user $u_i$ is given by $\sum_{k=1}^K v_k((\sum_{i=1}^k p_k)^{d(u_i)}-(\sum_{i=1}^{k-1} p_k)^{d(u_i)})$. Further, the expectation of (\ref{equ:newoptimal}) can be given by:
\begin{align}
\mathbb{E}_W[f^\star(G,W)]\leq \frac{1}{2} \sum_{u_i\in U}\sum_{k=1}^K v_k((\sum_{i=1}^k p_i)^{d(u_i)}-(\sum_{i=1}^{k-1} p_i)^{d(u_i)}). \label{equ:generaloptimal}
\end{align}
The proof is completed.

\section{Proof of Proposition 2}
Based on Proposition \ref{pro:generalupper}, we are able to compute the upper bound of the average total weight under the optimal matching for any given graph. We now consider a $n\times n$ grid and that the weight set $V=\{v_1=1,v_2=2\}$ and the weight probability distribution $P=\{p_1=\frac{1}{2},p_2=\frac{1}{2}\}$ are given. Note that, in this gird, there are $(n-2)^2$ nodes with degree $4$, $4n-8$ nodes with degree $3$, and $4$ nodes with degree $2$ in the grid. Moreover, the average maximum weight of a node with degree $d$ is $2-(\frac{1}{2})^d$. Thus, we finally obtain the average total weight under the optimal matching in the grid is upper bounded by: $\frac{31}{32}n^2-\frac{1}{8}n-\frac{1}{8}$.

Regarding the average total weight of greedy matching, we first note that for each $2\times n$ sub-grid, the greedy algorithm must add the vertical edges with weight $2$ between the two rows and the average number of such edges is given by $\frac{n}{2}$. Therefore, the average weight caused by adding the vertical edges with weight $2$ is equal to $n$. 

After adding these edges, the remaining graph becomes a lot of segments as shown in Fig.~\ref{fig:grid}. We first compute the average number of segments with length $t$ by using probability analysis, and it is given by $\frac{n}{2^{t+2}}$. Moreover, note that, for a segment with length $1$, the algorithm will add the only vertical edge with weight $1$. Therefore, the average weight caused by adding the vertical edges in segments of length $1$ is equal to $\frac{n}{2^{3}}$. 

As for for a segment with size $2\times t$ and $t>1$, its greedy matching consists of three parts: the individual greedy matching in the first row, the individual greedy matching in the second row, and all possible matching of the vertical edges between the remaining unmatched nodes of the two rows. Thus, the average total weight under the greedy matching in this segment should be at least equal to $2a_{t-1}$, where $a_{t-1}$ denotes the average total weight for a linear network with $t$ nodes. Therefore, the average weight caused by matched edges in segments of length $t>1$ is larger than or equal to $2\sum_{t=2}^{\infty}\frac{n}{2^{t+2}}a_{t-1}$. To compute that, we need the value of $a_t$ for any $t$. Note that $a_0=0$, $a_1=\frac{3}{2}$, $a_2=\frac{7}{4}$ and we have $a_t=\frac{7}{4}+\frac{3}{4}a_{t-2}+\frac{1}{4}a_{t-3}$ according to (\ref{equ:greedyrecurrence}). Thus, we can compute the value of $a_t$ for finite number of $t$, for example, from $t=1$ to $t=1000$. Then, we can further derive a lower bound for $\sum_{t=2}^{\infty}\frac{n}{2^{t+2}}a_{t-1}$, which is given by $\sum_{t=2}^{1000}\frac{n}{2^{t+2}}a_{t-1}\approx 0.26n$. Note that the value of $\sum_{t=1001}^{\infty}\frac{n}{2^{t+2}}a_{t-1}$ is almost zero since $2^{t+2}$ increases exponentially while $a_t$ increases linearly when $t$ is large.

In sum, we can prove that the average total weight under the greedy matching in the $n\times n$ grid is lower bounded by:
\begin{align}
\lim_{n\to \infty} \mathbb{E}_W [\hat{f}(G,W)]&> \frac{n}{2}(n+\frac{1}{8}n+2\sum_{t=2}^{1000}\frac{n}{2^{t+2}}a_{t-1})\nonumber\\
&> 0.8225 n^2,\nonumber
\end{align}
where $\mathbb{E}_W [\hat{f}(G,W)]$ denotes the average total weight under the greedy matching in the $n\times n$ grid and $a_{t}$ denotes the average total weight under the greedy matching in a linear network with $t$ edges.

Now, we know that the lower bound for the average total weight under the greedy matching is given by $\lim_{n\to \infty} \mathbb{E}_W [\hat{f}(G,W)]> 0.8225n^2$ and the upper bound of the average total weight under the optimal matching is given by $\frac{31}{32}n^2+o(n^2)$. Therefore, the average performance guarantee is $\frac{0.8225}{\frac{16}{32}}\approx 84.90\%$ when $n\to\infty$.

\section{Proof of Proposition 3}

We first consider the special case of $1\times n$ grid. For any user $u_i$ in such grid, let $I(u_i)$ be the indicator variable that the length $H(u_i)$ of the longest chain (sequence of edges) that has non-decreasing weights and starts from $u_i$ towards the left or right side is greater than $c\log n$ (c is a constant). Let $I(G)$ be the indicator variable that the linear graph $G$ has at least one such chain with length greater than $c\log n$. Then we have:

\begin{align}
\mathbb{E}(I(G))\leq\mathbb{E}(
\sum_{i=1}^n I(u_i))\leq 2nq, 
\end{align}
where $q$ denotes the probability that $c\log n$ consecutive edges has non-decreasing weights.

The weight of each edge is assumed to independently take value from $K$ kinds of weight values $\{v_1,v_2,\cdots,v_K\}$ according to the probability distribution $P=\{p_1,p_2,\cdots,p_K\}$. Then, for any $c\log n$ edges, there are totally:
\[\sum_{k=1}^{K-1}{{1+c\log n}\choose k}{K-2 \choose k-1},\]
kinds of non-decreasing weight combinations, and for each of the combinations, the probability to happen is upper bounded by $(\max\{p_1,p_2,\cdots,p_K\})^{c\log n}$. Therefore, the upper bound of the probability $q$ that any $c\log n$ consecutive edges have non-decreasing weights is given by:
\[q\leq (\max\{p_1,p_2,\cdots,p_K\})^{c\log n}\sum_{k=1}^{K-1} {1+c\log n \choose k}{K-2 \choose k-1}\]
\[\leq n^K (\max\{p_1,p_2,\cdots,p_K\})^{c\log n}.\]
Then, we have:
\[\mathbb{E}(I(G)) \leq 2n^{K+1}(\max\{p_1,p_2,\cdots,p_K\})^{c\log n}.\]
Note that when $c>\frac{K+1}{-\log \max\{p_1,p_2,\cdots,p_K\}}$, $\mathbb{E}(I(G))$ converges to $0$ when $n\to \infty$.

Therefore we can conclude that, in $1\times n$ grid, the algorithm run in $O(\log n)$ time w.h.p. according to the first moment method. 

Next, we extend our analysis to the $n\times n$ grid. Note that, in such grid, the number of possible chains that start from any given node $u_i$ and have non-decreasing weights is no longer two (toward left or right) as in $1\times n$ grid networks, but still limited to be less than $4K$ (the weight set size $K$ is a given constant). This is because we assume in Algorithm 1, every node will need to use priorities over ties among neighbors whose edge has the same weight and thus the number of possible chains that are relevant to a user's decisions is significantly reduced. Then we have:

\begin{align}
\mathbb{E}(I(G))\leq\mathbb{E}(
\sum_{i=1}^n I(u_i))\leq 4K n^2q, 
\end{align}
By using the similar arguments in the $1\times n$ grid, we can prove that in $n\times n$ grid, the algorithm run in $O(\log n)$ time w.h.p.. 

\section{Proof of Proposition 4}

In the random graph $G(n,p)$ with a large user number $n$ and a constant connection probability $p$, the probability that there exist $\sqrt{n}$ edges that have less edges than $\sqrt{n}$ among them is upper bounded by:
\begin{align}
&\lim_{n\to \infty} {n \choose \sqrt{n}}\sum_{i=0}^{\sqrt{n}-1} {\frac{\sqrt{n}(\sqrt{n}-1)}{2} \choose i} p^{i}(1-p)^{\frac{\sqrt{n}(\sqrt{n}-1)}{2}-i}\nonumber\\
&\leq \lim_{n\to \infty} (1-p)^{\frac{n-\sqrt{n}}{2}} {n \choose \sqrt{n}}^2\sum_{i=0}^{\sqrt{n}-1} (\frac{p}{1-p})^i= 0.\nonumber
\end{align}
Therefore, for any $\sqrt{n}$ nodes in $G(n,p)$, there are more than $\sqrt{n}$ edges among them with probability 1.

Let $E_i$ denote the number of edges in the random graph after the greedy matching algorithm adding $i$ edges. The probability that the heaviest edge among $E_i$ edges has weight $v_K$ is $1-(\sum_{k=1}^{K-1}p_k)^{E_i}$. Thus, the probability that the first heaviest edge to add in the greedy algorithm has weight $v_K$ is given by $1-(\sum_{k=1}^{K-1}p_k)^{E_0}$. After adding the first edge in the greedy algorithm, the probability that the heaviest edge to add among the remaining edges has weight $v_K$ is given by $(1-(\sum_{k=1}^{K-1}p_k)^{E_0})(1-(\sum_{k=1}^{K-1}p_k)^{E_1})\geq 1-(\sum_{k=1}^{K-1}p_k)^{E_0}-(\sum_{k=1}^{K-1}p_k)^{E_1}$. Similarly, we have that when $n\to \infty$, the probability that the $(n-\sqrt{n})/2$-th edge to add have weight $v_K$ satisfies:
\begin{align}
&\lim_{n\to \infty} 1-\sum_{i=0}^{(n-\sqrt{n})/2-1}(\sum_{k=1}^{K-1}p_k)^{E_i}\nonumber\\
&\geq \lim_{n\to \infty} 1-\frac{n-\sqrt{n}}{2}(\sum_{k=1}^{K-1}p_k)^{E_{(n-\sqrt{n})/2-1}}\nonumber\\
&\geq \lim_{n\to \infty} 1-\frac{n-\sqrt{n}}{2}(\sum_{k=1}^{K-1}p_k)^{\sqrt{n}}= 1,
\end{align}
where the second inequality is because the number $E_{(n-\sqrt{n})/2-1}$ of edges among the remaining $\sqrt{n}+2$ unmatched users is large than $\sqrt{n}$ with probability 1 when $n\to \infty$ as we have proved earlier, 

Therefore, we have that all the first $(n-\sqrt{n})/2$ edges added to the greedy matching have the largest weight $v_K$ with probability 1. Moreover, an obvious upper bound of the total weight under the optimal matching is given by $nv_K/2$ since any two users can be matched with a edge with weight $v_K$ at most. Therefore, the average performance guarantee of the greedy algorithm is given by:
\[\lim_{n\to \infty}\frac{v_K((n-\sqrt{n})/2)}{ v_K n/2}=1.\]
The proof is completed.

\section{Proof of Proposition 5}

Similarly, let $I(u_i)$ be the indicator variable that the length $H(u_i)$ of the longest chain (sequence of edges) that has non-decreasing weights and starts from $u_i$ towards the left or right side is greater than $c\log n$ (c is a constant). Let $I(G)$ be the indicator variable that the linear graph $G$ has at least one such chain with length greater than $c\log n$. Then we have

\begin{align}
\mathbb{E}(I(G))&\leq\mathbb{E}(
\sum_{i=1}^n I(u_i))=n \mathbb{E}(I(u_1))\nonumber\\
&\leq n(\frac{d(n-1)}{n})^{\log n}p< nd^{c\log n}p, \nonumber   
\end{align}
where $p$ denotes the probability that $c\log n$ consecutive edges has non-decreasing weights.

By using the similar arguments in the proof of Proposition 3, we have $p<n^K (\frac{\max\{p_1,p_2,\cdots,p_K\}}{2})^{c\log n}$. Then we obtain:
\[\mathbb{E}(I(G)) \leq (\frac{1}{2})^{-K} n^{K+1}d^{c\log n}(\frac{\max\{p_1,p_2,\cdots,p_K\}}{2})^{c\log n}.\]
\[=(\frac{1}{2})^{-K}n^{K+1-c(-\log \frac{\max\{p_1,p_2,\cdots,p_K\}}{2}-\log d)}\]
Note that when $d<\frac{2}{\max\{p_1,p_2,\cdots,p_K\}}$, there always exists a constant $c>\frac{K+1}{\log2-\log \max\{p_1,p_2,\cdots,p_K\}-\log d}$ that makes $\mathbb{E}(I(G))$ converge to $0$ when $n\to \infty$. Therefore we can conclude that the algorithm will terminate within $c\log n$ iterations w.h.p. according to the first moment method.

\section{Proof of Proposition \ref{pro:gnptree}}\label{app:gnptree}

In $G(n,d/n)$ with $d<1$, to prove \eqref{equ:gnptree}, we first show the connected component of any user $u_i$ has no loops w.h.p., and thus we can analyze the conditional expectation of $\mathbb{E}_{G\sim G(n,d/n)} [x_i(G)| C(u_i)=0]$ assuming the number $C(u_i)$ of loops in $u_i$'s component is zero, instead of directly analyzing $\mathbb{E}_{G\sim G(n,d/n)} [x_i(G)]$ (step 1 below). Then, it remains to show $\mathbb{E}_{G\sim G(n,d/n)} [x_i(G)| C(u_i)=0]$ can be well approximated by $\mathbb{E}_{T\sim T(d),W}[x_{root}(T,W)]$ in the approximated random tree $T(d)$, as both of them are considered in graphs without loops (step 2 below).

\textbf{Step 1: }
In $G(n,d/n)$, there are totally a random number of loops each of which includes at least three users, and for any $t\geq 3$ users, they have totally $\frac{(t-1)!}{2}$ kinds of permutations to form a loop. Thus, the average total number of loops is:
\begin{align}
\mathbb{E}(Loop)\hspace{-2pt}=\hspace{-2pt}\sum_{t=3}^n\hspace{-2pt}{n\choose t} \frac{(t\hspace{-2pt}-\hspace{-2pt}1)!}{2}(\frac{d}{n})^t\hspace{-2pt}<\hspace{-2pt}\frac{1}{6}\sum_{t=3}^n (d)^t\hspace{-2pt}<\hspace{-2pt}\frac{(d)^3}{6(1\hspace{-2pt}-\hspace{-2pt}d)},\label{equ:loopnumber}
\end{align}
which implies that regardless of the graph size $n$, there are at most a constant number of loops. Moreover, since \cite{gnpproperty} proves that $G(n,d/n)$ almost surely has no connected components of size larger than $c\log n$ for some constant $c$, the average size of the largest connected component in $G(n,d/n)$ is only $o(n)$. By combining this with \eqref{equ:loopnumber}, the average total number of users in the components with loops should be less than $o(n)\mathbb{E}(Loop)=o(n)$. Therefore, the probability that user $u_i$ is one of these users who are in components with loops is:
\begin{align}
\lim_{n\to \infty}Prob(C(u_i)\geq 1)=\frac{o(n)}{n}=0, \label{equ:probnocycle} \end{align}
where $C(u_i)$ denotes the number of loops in the connected component of user $u_i$. Based on \eqref{equ:probnocycle}, we now have:
\begin{align}
&\lim_{n\to \infty} \hspace{-2pt}\mathbb{E}_{G\sim G(n,\frac{d}{n})} [x_i(G)]\hspace{-2pt}=\hspace{-2pt}\lim_{n\to \infty}\hspace{-2pt} \mathbb{E}_{G\sim G(n,\frac{d}{n})} [x_i(G)| C(u_i)\hspace{-2pt}=\hspace{-2pt}0]. \label{equ:firstinequlity}
\end{align}

\textbf{Step 2: }
In this step, our problem becomes to prove that the conditional expectation $\mathbb{E}_{G\sim G(n,d/n)} [x_i(G)| C(u_i)=0]$ can be well approximated by $\mathbb{E}_{T\sim T(d),W}[x_{root}(T,W)]$. Given the connected component of user $u_i$ is a tree (i.e., with $C(u_i)=0$ loop), we start by first showing that the users in the component connect with each other in a similar way as in the random tree $T(d)$. 

We do a breadth-first search (BFS) starting from $u_i$ to explore all its connected users by searching all its direct neighbors prior to moving on to the two-hop neighbors. Note that the number of neighbors we explore from user $u_i$ follows the binomial distribution $B(n-1,d/n)$. In fact, for any user in the BFS tree of $u_i$, the number of new neighbors we directly explore from this user follows the binomial distribution $B(n-m,d/n)$ where $m$ is the number of users that have already been explored currently and cannot be explored again (otherwise a loop would occur). Meanwhile, in $T(d)$, each node gives birth to children according to the same Poisson distribution $Poi(d)$ no matter how large the tree currently is. 

Next, we prove that the difference between the branching under $B(n-m,d/n)$ and $Poi(d)$ is trivial. Actually, for any $t,m \leq c\log n$, the ratio of the probability of getting $t$ from distribution $B(n-m,d/n)$ and the probability of getting $t$ from distribution $Poi(d)$ is bounded by:
\begin{align}
1-\frac{1}{\sqrt{n}}\leq \frac {{n-m\choose t}(\frac{d}{n})^t(1-\frac{d}{n})^{n-m-t}}{e^{-d} d^t/(t)!}\leq 1+\frac{1}{\sqrt{n}},\label{equ:ratioboundpoi}
\end{align}
when $n$ is sufficiently large.

We define $\Theta$ as the set of all possible graph structures that the random tree $T(d)$ can have, and $p_1(\theta)$ as the corresponding probability for any structure $\theta\in \Theta$. As the set $\Theta$ includes all possible structures that the BFS tree of $u_i$ can have, we also define $p_2(\theta)$ similarly for the BFS tree. Then, based on \eqref{equ:ratioboundpoi}, for any $\theta\in \Theta$ with user size $s(\theta)< c\log n$, we have:
\begin{align}
(1-\frac{1}{\sqrt{n}})^{s(\theta)}\leq\frac{p_2(\theta)}{p_1(\theta)}\leq(1+\frac{1}{\sqrt{n}})^{s(\theta)},\label{equ:sizeratio}
\end{align}
which is because both the probabilities $p_1(\theta)$ and $p_2(\theta)$ are given by the product of all $s(\theta)$ users' individual probability to give birth to a given number of children as in structure $\theta$. 

We now note that the difference between $\mathbb{E}_{T\sim T(d)} [x_{root}(T)]$ and $\mathbb{E}_{G\sim G(n,d/n)} [x_i(G)| C(u_i)=0]$ is determined by the difference between $p_1(\theta)$ and $p_2(\theta)$:
\begin{align}
&\mathbb{E}_{T\sim T(d)} [x_{root}(T)]=\sum_{\theta\in\Theta}p_1(\theta) x_{root}(\theta),\label{equ:roottotal}\\
&\mathbb{E}_{G\sim G(n,d/n)} [x_i(G)| C(u_i)=0] =\sum_{\theta\in\Theta}p_2(\theta) x_{root}(\theta).\label{equ:gnptotal}
\end{align}
\eqref{equ:gnptotal} uses $x_{root}(\theta)$ instead of $x_{i}(\theta)$ because the matching weight $x_{i}(\theta)$ of node $u_i$ should be equal to the matching weight $x_{root}(\theta)$ of the root node when the BFS tree of $u_i$ has the same structure $\theta$ as the rooted tree. 

By substituting \eqref{equ:sizeratio} into \eqref{equ:roottotal} and \eqref{equ:gnptotal}, we have:
\begin{align}
&\lim_{n\to \infty} |\mathbb{E}_{T\sim T(d)} [x_{root}(T)]-\mathbb{E}_{G\sim G(n,d/n)} [x_i(G)| C(u_i)=0]|\nonumber\\
&=\lim_{n\to \infty} |\sum_{\theta\in\Theta} x_{root}(\theta)(p_1(\theta)-p_2(\theta)) | \nonumber\\
&\leq \lim_{n\to \infty} \frac{v_K}{2}\sum_{\{\theta\in\Theta:s(\theta)\geq c\log n\}} p_1(\theta)+ p_2(\theta) \nonumber\\
&+\frac{v_K}{2}\sum_{\{\theta\in\Theta:s(\theta)<c\log n\}}p_1(\theta)((1+\frac{1}{\sqrt{n}})^{s(\theta)}-1)\nonumber\\
&\overset{\text{(a)}}= \lim_{n\to \infty} \frac{v_K}{2}\frac{1}{\sqrt{n}} \sum_{\{\theta\in\Theta:s(\theta)<c\log n\}} p_1(\theta)s(\theta)\nonumber\\
&\overset{\text{(b)}}\leq \lim_{n\to \infty} \frac{v_K}{2} \frac{1}{\sqrt{n}} \frac{1}{1-d} =0, \label{equ:secondtinequlity}
\end{align}
where we split the analysis in two cases depending on whether the graph structure $\theta$ has size larger than $c\log n$ or not. As we mentioned earlier, \cite{gnpproperty} proves that $G(n,d/n)$ almost surely has no connected components of size larger than $c\log n$ for the constant $c$, thus we have the probability $\sum_{\{\theta\in\Theta:s(\theta)\geq c\log n\}} p_2(\theta)=0$ to derive equality (a). Inequality (b) is due to that the average size of the random tree $T(d)$ is given by $\sum_{\theta\in\Theta} p_1(\theta)s(\theta)=\sum_{t=0}^\infty d^t=\frac{1}{1-d}$ where $d^t$ is the average size of the $t$-th generation as each node gives birth to $d$ children on average. Moreover, according to the first moment method, we also have $\sum_{\{\theta\in\Theta:s(\theta)\geq c\log n\}} p_1(\theta)=0$ for (a). 

Based on \eqref{equ:firstinequlity} and \eqref{equ:secondtinequlity}, we finally prove \eqref{equ:gnptree}.

\section{Proof of Proposition \ref{pro:gnptreeyk}}\label{app:gnptreeyk}

To compute the proposal probability $y_k$, we further define $y_k^c$ to denote the probability that the root node receive a proposal from a child who connects to it with an edge of weight $v_k$ given this child gives birth to $c$ grandchild nodes (happens with probability ${n-1 \choose c}(d/n)^c(1-d/n)^{n-1-c}\to\frac{d^c}{c!}e^{-d}$ as $n\to \infty$). If the edge between the root node and the child has the maximum weight (i.e., $k=K$), the child will send a proposal to the root node only when all $i$ grandchildren that have the maximum weight among the $c$ grandchild nodes (happens with probability ${c \choose i}(p_K)^i (1-p_K)^{c-i}$) either want great-grandchildren (happens with probability $1-y_K$) or have lower priority to be added. Thus, the recursive equation for the proposal probability $y_K^c$ is given as follows:
\[y_K^c=\sum_{i=0}^c {c \choose i}(p_K)^i (1-p_K)^{c-i}(\sum_{j=0}^i {i\choose j}y_K^{i-j}(1-y_K)^j\frac{1}{i-j+1}).\]
Moreover, by considering all possible number of grandchildren that the child can give birth to, we can derive the aggregate recursive equation for the matching possibility $y_K$:
\begin{align}
y_K&=\sum_{c=0}^{\infty} \frac{d^c}{c!}e^{-d} y_K^c
\nonumber\\ 
&=\sum_{c=0}^{\infty} \frac{d^c}{c!}e^{-d} \sum_{i=0}^c {c \choose i}(p_K)^i (1-p_K)^{c-i}(\sum_{j=0}^i \frac{{i\choose j}y_K^{i-j}(1-y_K)^j}{i-j+1}) \nonumber\\ 
&=e^{-p_K d}\sum_{i=0}^\infty  (\frac{(p_Kd)^i(1-(1-y_K)^{i+1})}{(i+1)!y_K},\label{equ:first}
\end{align}
after a summation by parts. Note that the term $\frac{1}{y_K}(1-(1-y_K)^{i+1})$ on the right-hand-side of the equation is decreasing in $y_K$ when $y_K\in (0,1)$. Moreover, when $y_K=0$, the RHS of the equation is equal to 1. When $y_K=1$, the RHS of the equation is equal to $\frac{1}{p_Kd}(1-e^{-p_Kd})<1$ for any $d>0$ and $0<p_K<1$. Therefore, there exist a unique solution $y_K^\star$ satisfying the above equation in the interval $(0,1)$.

Then, after the proposal probability $y_{K}$ of the maximum weight has been decided, the proposal probability $y_{K-1}$ now have the highest priority to compute as $w_{K-1}$ becomes the maximum weight among the remaining weights. Similarly, for the proposal probability $y_{K-1}$, we can derive the following recursive equation:
\begin{align}
y_{k}=e^{-(p_K+ y_Kp_K) d}\sum_{i=0}^\infty \frac{(p_kd)^i(1-(1-y_k)^{i+1})}{(i+1)!y_k}.\label{equ:second}
\end{align}
Using the similar argument for $y_{K}$, we prove that there exist a unique solution $y_{K-1}^\star$ satisfying the above equation in the interval $(0,1)$ after substituting the solution $y_K^\star$ to (\ref{equ:first}) into (\ref{equ:second}). 

Eventually, for any $k=1,2,\cdots K$, we can derive the following recursive equation:
\begin{align}
y_{k}=e^{-(p_K+\sum_{j=k+1}^K y_jp_j) d}\sum_{i=0}^\infty \frac{(p_kd)^i(1-(1-y_k)^{i+1})}{(i+1)!y_k}.\nonumber
\end{align}
and prove that there exists a unique solution to the equation.

\end{appendices}

\end{document}